\documentclass{article}
\pdfoutput=1
\usepackage{arxiv}

\usepackage[utf8]{inputenc} 
\usepackage[T1]{fontenc}    
\usepackage{hyperref}       
\usepackage{url}            
\usepackage{booktabs}       
\usepackage{amsfonts}       
\usepackage{nicefrac}       
\usepackage{microtype}      
\usepackage{lipsum}
\usepackage{biblatex}
\usepackage{graphicx}       
\usepackage{multicol}       
\usepackage{multirow}       
\usepackage{pdflscape}      

\title{Locked-in Syndrome Machine Learning Classification using Sentence Comprehension ERP EEG Data}

\author{
Daniël van den Corput\\
Department of Cognitive Science \& Artificial Intelligence\\
Tilburg University\\
\texttt{d.p.t.vdncorput@tilburguniversity.edu}
\\
}

\begin{document}
\maketitle

\begin{abstract}
Locked-in Syndrome patients are often misdiagnosed and face pessimistic prognosis because of similarities with disorders of consciousness, a lack of objective biomarkers and a difficult-to-recognize pathogenesis. Biomarkers show promise in identifying similar conditions, utilizing electroencephalography (EEG) data. This data, particularly in the form of event-related potentials (ERPs), while successful in varying applications, suffers from methodological constraints and interpretation obstacles. The study documented in this body of work explores a machine learning paradigm with regards to N400 ERP data retrieved from a sentence comprehension task to tackle these hindrances and proposes a new auxiliary diagnostic tool for LIS and possibly disorders of consciousness. A support vector machine (SVC) and a random forest classifier (RF) were able to classify conscious individuals from unconscious ones with optimistic performance metrics. Based on these results, the proposed models and continuations thereof present valuable opportunities for the development of an auxiliary diagnostic tool for the classification of LIS patients, aiding diagnosis, improving prognosis, stimulating recovery and reducing mortality rates.
\end{abstract}


\section{Introduction}
\label{Introduction}
Among coma and coma-like states, Locked-in Syndrome (LIS) might be one of the hardest diagnostic challenges medical professionals are facing today. A complete description of LIS, contemporary diagnostic challenges and attempts made to overcome these challenges can be found in Appendix \ref{appendix:bginfo}. Even though recent systematic research regarding the prognosis of LIS patients is very scarce, early research found mortality rates as high as 60\% \cite{patterson1986locked}. This is unsettling since early diagnosis, effective care and appropriate rehabilitation reduce the mortality rate significantly and allow for cognitive and motor function recovery, verbal communication and independent breathing patterns \cite{casanova2003locked}. The disconcerting truth, however, is that LIS is one of the most likely disorders to result in misdiagnosis for multiple reasons. Therefore, identifying objective and specific assessment markers is paramount to enable early diagnosis. This is a pivotal goal throughout psychiatry where a lack of specific and objective assessment markers is prevalent \cite{takizawa2014neuroimaging}.

\par Assessing consciousness is a necessity when differentiating between various disorders of consciousness (DOC), which is imperative given that particular disorders require their own approaches and therapeutic decisions, influencing prognosis \cite{noirhomme2017look}. Suggestively, the current gold standard in assessing consciousness in LIS-patients is based on bedside behavioral examinations (BBEs), despite 40\% of these assessments resulting in misdiagnosis \cite{gosseries2014recent}. Even though LIS is not a DOC, many characteristics are shared among the two (see Appendix \ref{appendix:bginfo}), resulting in regular confusion of the former for the latter by virtue of diagnostic hindrances \cite{bruno2011unresponsive}.

\par The lack of adequate assessment markers emanates from the complex pathogenesis of psychiatric and neurological conditions, inherently linked to human cognition and behavior \cite{lui2016psychoradiology}. Biomarkers have been proposed as a main contender to overcome this issue, given their ability to improve diagnosis, beneficially impact prognosis and tailor individual treatments \cite{singh2009biomarkers}. Functional imaging has proven successful because of its capability to identify specific disorders and to differentiate between similar disorders based on these biomarkers \cite{takizawa2014neuroimaging}. However, imaging tools are often expensive and distressing for patients to use, suffer from methodological constraints and are open to subjective interpretability \cite{takizawa2014neuroimaging,gosseries2014recent, noirhomme2017look}. Because it can be collected non-invasively, is easy to apply and relatively low in cost, EEG data is the most popular imaging tool and used widely for various applications \cite{mcloughlin2014search,thul2016eeg}. Current EEG advancements allow for both high spatial and temporal resolution, increasing its allure for biomarker identification \cite{mcloughlin2014search}. Simultaneously, interpreting EEG signals still poses challenges because of its high inter-rater variability and difficulty of interpretation \cite{acharya2018deep}. A machine learning (ML) approach has been suggested to overcome these issues due to the fast, consistent, accurate and relatively objective diagnoses it can achieve \cite{acharya2018deep}. Event-related potentials (ERPs) in particular have been identified as important clinical and research instruments in various tasks and modalities \cite{kutas1983brain,vecchio2011auditory,steppacher2013n400,balconi2013disorders,beukema2016hierarchy,rohaut2015probing}. ERPs are portions of EEG recordings that directly reflect cortical neuronal activity following particular events \cite{jung1999analyzing} (a more detailed explanation can be found in Section \ref{RelatedWork}).

\section{Related Work}
\label{RelatedWork}
Since its invention in 1924, EEG signals were not used to make an attempt to break down the concept of consciousness until 1961 \cite{henrie1961alteration}. Since then, EEG research has made use of aspects like bandwidth, spectral data and power spectra to make classifications of some kind, but generally lacked clinical relevance because of processing difficulties, sources of errors and a lack of adequate classification techniques \cite{walter1967discriminating,larsen1970automatic,struys1998comparison}. Even though difficulty of interpreting is still a relevant challenge within EEG processing, the emergence of ML has increased its relevance as a clinical tool \cite{acharya2018deep,fergus2016machine}. Advancements in processing techniques have, among other things, lead to systems that can automatically remove artifacts and extract features \cite{winkler2011automatic,alomari2013automated}, increasing the number of studies in which ML is used for various diagnostic purposes. EEG in combination with ML has, for example, been widely used to recognize and classify Alzheimer’s disease by means of principal component linear discriminant analysis, bagging, random forests, support vector machines and feed-forward neural networks \cite{lehmann2007application,trambaiolli2011improving,liu2014early}. In these cases, random forests, support vector machines and neural networks achieved the highest sensitivity and specificity, up to 89\% and 87\% respectively, even for mild patients \cite{lehmann2007application}. ML has also been effective in the recognition and classification of different states of consciousness. Techniques like SVM, k-nearest neighbor and linear discriminant analysis were used again to successfully classify patients with DOC \cite{holler2013comparison,engemann2018robust}. The common conclusion in these studies is that ML in combination with EEG markers of consciousness are effective for diagnosis, recognition and discrimination in various clinical contexts because they are reliable, low in cost, automatic and fast.

\par ERP's were first applied to the context of language by 
\citeauthor{kutas1983brain} \cite{kutas1983brain}, who studied the effect of grammatical errors and semantic anomalies on elicited ERPs. They found that subjects who were presented with semantically anomalous words and grammatical errors elicited a distinct N400 ERP effect in numerous scalp regions, most prominently for unexpected words within their context \cite{kutas1983brain}. The N400 effect is described as the negative ERP peak following 300-500ms after experimental onset \cite{kutas2011thirty}. They have been found in a range of different stimulus types in various language processing tasks, applied to detect many aspects of language and established as a valid indicator for the surprisal value of a word in its presented context \cite{kutas1983brain,kutas2011thirty}. This surprisal value is represented by a Cloze score, where scores are low, medium and high for 0\%-33\%, 34\%-66\% and 67\%-100\%, respectively \cite{bormuth1968cloze}. This effect was experimentally confirmed by \citeauthor{nicenboim2020words} \cite{nicenboim2020words}, both for items with a low Cloze score (contstraining) and a high Cloze score (non-constraining). Items with a high Cloze score or in a non-constraining context elicited a more negative ERP peak about 300-500ms after onset as compared to items with a low Cloze score or in a constraining context in their study.

\par These ERP effects have moreover been used to detect and measure consciousness in various experimental settings. \citeauthor{steppacher2013n400} \cite{steppacher2013n400} used ERPs elicited by sound (P300) and speech (N400) to measure consciousness by assessing information processing in unresponsive wakefulness syndrome- and minimally conscious patients. They identified N400 effects in 32\% of unresponsive wakefulness syndrome patients and 41\% of minimally conscious syndrome patients. Specifically the N400 effect elicited by non-constraining words was identified to predict long term recovery and prognosis in these patients, but was not able to differentiate between the two. Stronger elicited N400 effects predicted a more favourable clinical outcome. \citeauthor{balconi2013disorders} \cite{balconi2013disorders} used the semantic anomalies described by \citeauthor{kutas1983brain} \cite{kutas1983brain} to detect N400 peaks to verify preservation of linguistic processing in DOC and minimal-consciousness states to determine the level of consciousness among them. Even though they concluded that the found differences within their experiment were not enough to differentiate between different DOC, differences in N400 latency were able to differentiate between control subjects and DOC patients. \citeauthor{beukema2016hierarchy} \cite{beukema2016hierarchy} measured  N400 effects with regards to auditory stimuli in patients with DOC. Even though they could not significantly differentiate between DOC, they did identify ERPs following auditory stimuli as a possible clinical tool to improve accuracy of diagnosis and prognosis of DOC patients. In their experiment, 44\% of patients showed markers of N400 processing of speech and noise. To transform this process to be used as a valid clinical tool, sensitivity has to be improved first. This conclusion was also made by \citeauthor{rohaut2015probing} \cite{rohaut2015probing}, who probed semantic processing to determine measures of consciousness in non-communicating patients. They used N400 effects and late positive components effectively to differentiate between conscious, minimally conscious and vegetative state individuals, but emphasised that these effects have to be found easily and robustly at an individual level in order to develop a valid clinical diagnostic tool. \citeauthor{cruse2014reliability} \cite{cruse2014reliability} compared the effect of different stimuli and task demands on N400 amplitudes among healthy subjects, and detected 50\% N400 effects among subjects instructed to passively pay attention to normatively associated word-pairs.

\par In summary, ERP and N400 effects in particular have been identified to be capable of differentiating healthy subjects from patients with various DOC, unresponsive wakefulness syndrome or minimal consciousness, but generally lack clinical relevance because of low sensitivity or a lack of significant single-subject level differences. Moreover, N400 ERP effects with regards to Cloze scores offer the possibility to establish linguistic functions and as such detect consciousness among potential LIS patients. The fact that words can be presented passively \cite{steppacher2013n400,balconi2013disorders,beukema2016hierarchy,rohaut2015probing,cruse2014reliability} makes data collection convenient and increases the potential of N400 ML classification as the basis as diagnostic tool. Detection rates found in these studies (32\% and 41\% in \citeauthor{steppacher2013n400} \cite{steppacher2013n400}, 50\% in \citeauthor{cruse2014reliability}
\cite{cruse2014reliability}) are expected to be surpassed, while the sensitivity rate of 89\% found by \citeauthor{lehmann2007application} \cite{lehmann2007application} is aspired in the present study, so that the basis for an auxiliary LIS diagnostic tool is created.

\section{Objective}
\label{Objective}
Given the scenario presented in Section \ref{RelatedWork} with the additional information in Appendix \ref{appendix:bginfo}, the present study explores a ML-driven approach to classify conscious LIS patients from unconscious individuals based on EEG N400 ERP effects.

\section{Methods}
\label{Methods}
\subsection{Experimental Setup} \label{ExperimentalSetup}
In order to do so, data from \citeauthor{nicenboim2020words} \cite{nicenboim2020words} will be explored, extrapolated and adjusted. This data revolves around predictability effects of sentential context during a sentence comprehension task. Principally unrelated to LIS, the dataset will serve as the basis for the simulation of unconscious patients. Simulating EEG data has been successfully done throughout various ML studies, where different techniques propose particular merits and demerits \cite{haufe2013critical,yao2005evaluation,owen2012performance, delorme2007enhanced,haufe2016simulation,moosmann2008joint}. In some experimental setups simulated data is preferred over real data, because the inherently unsupervised nature of the latter entails an absence of ground truth values \cite{haufe2013critical}. In simulated data, the exact source locations are known, which is a great advantage because it allows for testing performance on challenging source configurations \cite{yao2005evaluation,owen2012performance}, generally problematic in natural EEG data \cite{delorme2007enhanced}. It also allows for objective and convenient testing of different methods, signal-to-noise ratios and dynamic characteristics \cite{haufe2016simulation}. On the other hand, it is important to accentuate that simulated data is essentially unrealistic, because it does not include any electrode measurement errors \cite{yao2005evaluation}, it remains difficult to define valid performance measures \cite{haufe2016simulation} and simulation is often done based on a Gaussian assumption, which generally represents natural data, but does not completely fit all natural scalp distributions \cite{moosmann2008joint}. Generated data should therefore have as much in common with the original data as possible, but should differ solely in the property being scrutinized in the study at hand \cite{haufe2013critical}.

\par The lion’s share of these simulations aim to generate signals from conscious individuals, while much less work is done to simulate signals of unconscious individuals. Given the lack of a theoretical foundation, unconscious patients were simulated based on the context of the original data. Since this data represents individuals being able to sensibly process words and thus representing conscious labels in the scope of this study, unconscious labels were yet to be acquired. Sensibility of conscious participants is reflected by the relationship between their EEG signals and the corresponding Cloze score of their experimental onset. It is expected that for conscious individuals, higher Cloze scores will elicit larger negative N400 peaks given the earlier described negative relationship between the two (see Section \ref{RelatedWork}). This relationship served as the basis for the simulation of unconscious individuals, for which EEG and ERP signals were forged and paired with randomly generated Cloze scores. By doing so, the inverse relationship true for conscious individuals will not be present in simulated individuals, assuming them to be unconscious. Since this simulation operates from the basis of Cloze scores rather than EEG data, methodological constraints commonly faced during simulation were circumvented. It also ensured that the simulated data deviated solely on the features of interest to ensure validity of the model.

\par When simulation was achieved, a support vector machine (SVM) and a random forest classifier (RF) were trained on preprocessed EEG signals. These models were subsequently used to explore to what extent conscious LIS patients can be distinguished from unconscious patients. The goal of doing so is to establish a foundation for the automatic diagnostic tool used to support traditional diagnostic methods in the assessment of LIS patients. Effective algorithms should be capable of measuring faint cortical signals inherent to LIS patients and as such reduce diagnosis time, ameliorate diagnostic procedures and bolster early diagnosis to ensure greater confidence in the assessment and treatment of LIS patients to ultimately reduce mortality rates. ML has been suggested, explored and applied as a solution for numerous diagnostic obstacles faced in functional imaging studies, but more research is necessary to ensure adequate appliance in the diagnostic process of LIS. The present study therefore built on this foundation of work with a model specifically designed to classify conscious LIS patients from unconscious patients based on EEG data.

\subsection{Participants}
\label{Paticipants}
From the 120 original subjects sampled by \citeauthor{nicenboim2020words} \cite{nicenboim2020words}, 110 subjects were available in the database at the moment of retrieval. EEG data from these 110 subjects was retrieved from the OSF database \cite{OSFwebsite}. This data served as the basis for the simulation of unconscious patients. The age of all these participants was between 18 and 35 years old and represented both a university subject-pool and community-based population. This number of participant is much larger than in typical ERP studies (N=30; \citeauthor{nicenboim2020words} \cite{nicenboim2020words} as well as typical studies regarding LIS classification (2 <= N <= 23 in 
\citeauthor{noirhomme2017look}  \cite{noirhomme2017look} and classification of DOC (for example 25, 23 and 14 in  \cite{zheng2017disentangling,owen2012performance,wielek2018sleep}, respectively).

\subsection{Data} \label{Data}
The kick-off of this study was to further transform the preprocessed data from \citeauthor{nicenboim2020words}'s experiment, retrieved from \cite{nicenboim2019eegdata}, in order for them to be suitable for a ML classification algorithm. A description of their dataset and their preprocessing workflow can be found in Appendix \ref{appendix:originaldata}. Their data was used to represent conscious individuals on the assumption that they were capable of processing sentences sensibly, independent of their performance.

\par This dataset, from here on referred to as original data, consisted of individual .RDS files for every subject. Every file included EEG data points collected over time, information about experimental events and summarized information about the collected data. The code used in this study can be consulted as Github repository, in \cite{githubrepo}, and is explained below.

\subsection{Design \& Procedure} \label{DesignProcedure}
The original files were looped through and preprocessed individually. A list of the used software and packages to do so can be found in the footnote\footnote{\label{note1}Complete list of software: R (RStudio Team, 2019), and the R packages \textit{readr} (Version 1.3.1; Hester \& Francois, 2018), \textit{stringr} (Version 1.4.0; Wickham, 2019), \textit{eeguana} (Nicenboim, 2018), \textit{osfr} (Wolen, et al., 2020), \textit{caret} (Kuhn, 2020), \textit{dplyr} (Version 0.8.5; Wickham, François, Henry \& Müller, 2020), \textit{purrr} (Version 0.3.3; Henry \& Wickham, 2019), \textit{tidyr} (Version 1.0.2; Wickham \& Henry, 2020), \textit{tibble} (Version 3.0.1; Müller \& Wickham, 2020), \textit{factoextra} (Version 1.0.7; Kassambara \& Mundt, 2020), Jupyter Notebook (Kluyver, et al., 2016), and the Python packages \textit{NumPy} (Oliphant, 2006), \textit{Pandas} (McKinney, et al., 2010), \textit{Matplotlib} (Hunter, 2007), \textit{Scikit-learn} (Pedregosa, et al., 2011) and \textit{MNE-C} (Gramfort, et al., 2014).}. Below follows the demarcation of the created pipeline, displayed in Appendix \ref{Appendix3}. 

\par
\begin{figure}
    \centering
    \includegraphics[scale=0.70]{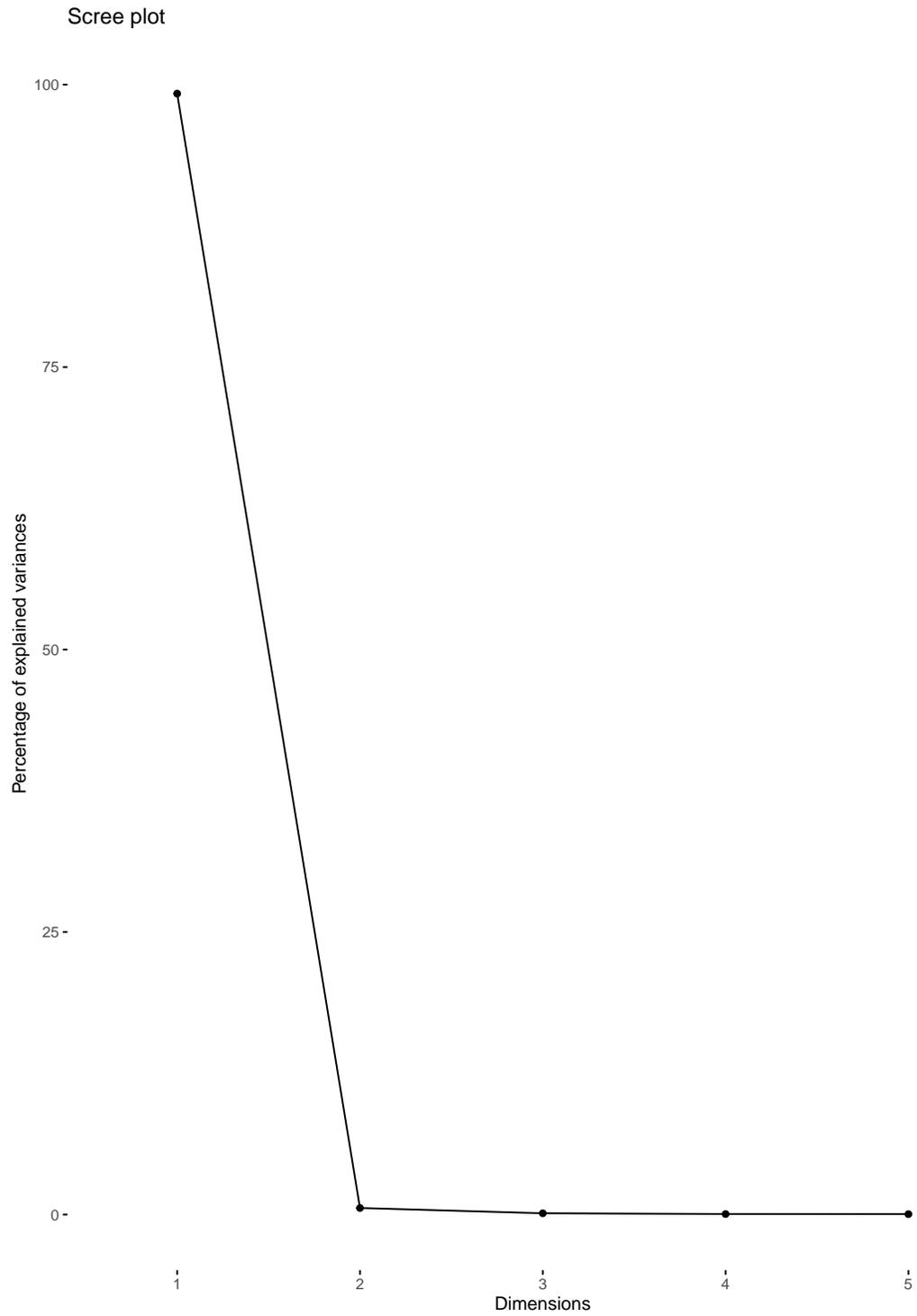}
    \caption{Scree plot of randomly picked subject. See \url{https://github.com/DanielvdC/LISclassification/blob/master/Visualizations/Scree\%20plot.pdf} for full size}
    \label{fig:pca}
\end{figure}

 Many of the features present in the original files were irrelevant for the study at hand, which is why they were discarded. First of all, channels related to the N400 ERP were defined and extracted from all the available recorded EEG channels. The N400 effect has been established as a negativity with a centroparietal distribution in the Cz, CP1, CP2, P3, Pz, P4 and POz channels within a time window of 300-500ms after experimental onset \cite{kutas2011thirty}. Neuronal signals are most anticipated within these particular channels because they are the most susceptible to experimental manipulation in this particular timeframe \cite{delong2014predictability}. Every datafile was further reduced to the seven N400 ERP channels as described above, the Cloze score of the corresponding noun and the feature \textit{constraint}. Next, signals from the ERP channels were downsampled by factor four, from the original 512 Hz to 128 Hz. This factor was chosen because it reduces the data to the greatest extent while still fitting the Shannon-Nyquist sampling theorem \cite{shannon1949communication} requirement, which states that a signal can be replaced by a discrete sequence of signals without losing any information if this signal has a sampling frequency of at least double the bandwidth (i.e. the Nyquist rate) \cite{jerri1977shannon}. Next, signals were binned with a small margin to the onset of the N400 ERP, ranging from 250 to 600ms after the onset of the experimental determiner. Data was filtered on items that were constraining, which are items presented in a context with nouns with a low Cloze score. The primary aim is to find whether a ML algorithm can classify conscious LIS patients from unconscious patients based on their EEG data. It was deemed appropriate to focus on constraining items so that data from the most generic sentences will be analyzed. By doing so, the model will be able to make predictions based on EEG data from natural reading tasks, warranting generalizability of the diagnostic tool. Since EEG data is of a rich format, a principal component analysis (PCA) was performed in order to reduce features and to decrease training and prediction time. The PCA algorithm performs a factor analysis to reduce data dimensionality by transforming features in such a way that they capture as much variation present in the original data as possible. These components have been established as valid input for classification algorithms, and subsequently used in further analysis \cite{wang2010feature}. As such, the PCA was performed on the ERP electrodes so that these were reduced to their principal components. The cross-validation method composed by \cite{wold1978cross} was applied, so that five, two and one principal components were all taken into the hyperparameter tuning process. It was assumed that two and one principal components would be enough to generate high performance metrics based on three extracted eigenvalues of randomly selected subjects, for which the values of the first component were 99.2\%, 99.4\% and 99.2\% respectively. This is indicated for one of these subjects in Figure \ref{fig:pca}. The original seven electrode ERP channels were nevertheless iteratively replaced with five, two and one components to assure best performance.

Subsequently, the .csv file was imported into Python for the final preprocessing steps and to set up the classification model because of the abundance of convenient processing tools. Necessary packages\footnote{See footnote 1.} were imported, the .csv file was loaded in and duplicated in order to generate data for unconscious individuals. The simulation was performed by means of a random uniform distribution between the minimum and maximum Cloze values of the original data. These generated numbers were then shuffled and inserted at the end of the duplicated data frame, upon which it was concatenated with the original one. This complete dataset was transformed so that each subject was represented by one row with all their corresponding principal components and Cloze scores as feature. As such, the dataframe now consisted of 220 participants with even conscious and unconscious classes, which were finalized by adding conscious labels (See Appendix \ref{Appendix3} for the complete pipeline).

\subsection{Models} \label{models}
The supervised classification was performed on a single-subject level, in which principal components (representing EEG signals) and Cloze scores were used to train and validate two classifiers, an SVM and an RF. 

\par Before splitting the data in a train- and test-set, missing values were filled, labels were one-hot encoded where the conscious class was represented by 1 and the unconscious by 0, the train and test size as well as the number of folds for the cross validation were set. Even though decision trees and RFs generally handle missing values within their algorithm, SVMs do not and are sensitive to them \cite{garcia2010pattern}. Therefore, missing values were filled by means of a linear interpolation with the pandas module \cite{mckinney2010data}, so that index was ignored and values were treated as equally spaced. This ensures that missing data was estimated based on a linear distribution, which most closely represents natural EEG data \cite{moosmann2008joint}. Since linear classifiers, like the linear kernel of the SVM, depend on the inner products of feature vectors \cite{sartakhti2012hepatitis}, data was standardized using the scikit-learn StandardScaler module \cite{pedregosa2011scikit}, which standardizes features by removing the mean and scaling to unit variance. The most basic SVM and RF were set up using scikit-learn modules \cite{pedregosa2011scikit}, which were thrown into a grid search to determine optimum parameters. In the case of the SVM this concerned the kernel, C and $\gamma$ parameters, where kernel represents the used kernel, C the tradeoff of correct classification versus maximization of the decision function’s margin and $\gamma$ how far the influence of a single training example reaches. Low values encourage a larger margin and far influence, high values encourage a smaller margin and close influence, respectively \cite{pedregosa2011scikit}. For the RF, this concerned the maximum depth and features, minimum samples required to be at a leaf node and to split a node and the number of estimators to use. An overview of the considered and actually implemented (hyper)parameters of these algorithms can be found in Table \ref{tab:gridsearch}, in which the best (hyper)parameters are presented in bold. Both algorithms were updated with the best parameters provided by the grid search. Next, training and validation sets were derived by means of fivefold cross validation, against which accuracy metrics were checked. For training and validation purposes, evaluation metrics were confined to accuracy, sensitivity and specificity. The test set was also evaluated with the receiver operating characteristic (ROC) curve. Hyperparameter tuning consisted of changing the train-test ratio, the number of principal components and the number of folds to use in the cross-validation.

\par An additional analysis was performed to test the model for its robustness. In order to do so, the original dataset was reduced to just over half its size by filtering on a subset of regions of the presented sentences. These sentences originally consisted of the equally distributed regions adjectives, determiners, nouns, pre-critical and post-critical, while this was reduced to the regions nouns, pre-critical and post-critical for the robustness check. N400 effects are produced as a response to the presented adjective in combination with its presented noun. However, the effect itself is only anticipated 250-600ms after this onset, possibly enabling the model to be trained on just the latter three regions without losing any predictive power.

\begin{table}
    \caption{Grid search of models, considered (hyper)parameters and \textbf{best} (hyper)parameters.}
    \label{tab:gridsearch}
    \newlength\q
    \setlength\q{\dimexpr .3\textwidth -2\tabcolsep}
    \noindent\begin{tabular}{p{\q}p{\q}p{\q}}
    
        \toprule
        Model                   &   (Hyper)Parameter    &   Values \\
        \midrule
        \multirow{4}{4em}{SVM}  &   Kernel              &   [\textbf{linear}, rbf] \\
                                &   C                   &   [\textbf{1}, 10, 100]  \\
                                &   $\gamma$             &   [\textbf{0.1}, 0.01,                                         0.001, 0.0001]  \\
                                &   PCA                 &   [\textbf{1},2,5]        \\
        \midrule
        \multirow{6}{6em}{RF}   &   Maximum depth       &   [\textbf{80}, 90, 100,                                                                          110] \\
                                &   Maximum features    &   [\textbf{2}, 3]      \\
                                &   Min. samples leaf   &   [\textbf{3}, 4, 5]  \\
                                &   Min. samples split  &   [\textbf{8}, 10, 12]    \\
                                &   No. of estimators   &   [\textbf{100}, 200, 300,                                                 1000] \\
                                &   PCA                 &   [1, \textbf{2}, 5]      \\
        \bottomrule
 \end{tabular}
 \end{table}
 
\section{Results}
The workflow was implemented on a workstation with a 1.4 GHz Dual-Core Intel Core i5 processor and 8 GB 1600 MHz random-access memory. The grid search with cross validation took about 25 minutes to complete, upon which the SVM and RF were set up and trained in about 15 seconds each. Prediction on the test set took about 2 seconds for each model.
\par Table \ref{tab:results} shows the confusion matrix and performance metrics for both models when predicting on the test set averaged over all five folds based on the best (hyper)parameters attained by predicting on the validation sets. The SVM scored 78\%, 80\%, 72\%, 94\% on area under the curve, accuracy, sensitivity and specificity respectively, where the RF scored 91\% on all four. The table also shows that the SVM produced more false positives, but one less false negative.
\par The robustness check indicates that the models were both robust to the decrease of input data. Both models produced very similar performance metrics, where the SVM produced one more false positive and the RF refrained from producing false negatives.

\begin{landscape}
    \thispagestyle{empty}
\begin{table}
    \caption{Confusion matrix and performance metrics of considered models.}
    \label{tab:results}
    \setlength\q{\dimexpr .18\textwidth-2\tabcolsep}
    \renewcommand{\arraystretch}{1.5}
    \noindent\begin{tabular}{p{\q}p{\q}p{\q}p{\q}p{\q}p{\q}p{\q}p{\q}}
    
        \toprule
                    &   \multicolumn{2}{l}{Predicted}   &   &   AUC     &   Accuracy    &   Sensitivity   &   Specificity \\
                    
                    &   &   Conscious   &   Unconscious \\ 
        \midrule
        \multirow{1}{*}{SVM} \\
        \midrule
        \multirow{2}{*}{True}   &   Conscious   &   21  &   1   &   \multirow{2}{*}{.78}    &   \multirow{2}{*}{.80}  &   \multirow{2}{*}{.72}    &   \multirow{2}{*}{.94} \\
        &   Unconscious &   8   &   15  \\
        \midrule
        \multirow{1}{*}{RF} \\
        \midrule
        \multirow{2}{*}{True}   &   Conscious   &   20  &   2   &   \multirow{2}{*}{.91}    &   \multirow{2}{*}{.91}  &   \multirow{2}{*}{.91}    &   \multirow{2}{*}{.91} \\
        &   Unconscious &   2   &   21  \\
        \bottomrule
        \textbf{Robustness check} \\
        \toprule
                    &   \multicolumn{2}{l}{Predicted}   &   &   AUC     &   Accuracy    &   Sensitivity   &   Specificity \\
                    
                    &   &   Conscious   &   Unconscious \\ 
        \midrule
        \multirow{1}{*}{SVM} \\
        \midrule
        \multirow{2}{*}{True}   &   Conscious   &   21  &   1   &   \multirow{2}{*}{.78}    &   \multirow{2}{*}{.78}  &   \multirow{2}{*}{.70}    &   \multirow{2}{*}{.93} \\
        &   Unconscious &   9   &   14  \\
        \midrule
        \multirow{1}{*}{RF} \\
        \midrule
        \multirow{2}{*}{True}   &   Conscious   &   22  &   0   &   \multirow{2}{*}{.96}    &   \multirow{2}{*}{.96}  &   \multirow{2}{*}{.92}    &   \multirow{2}{*}{1} \\
        &   Unconscious &   2   &   21  \\
        \bottomrule
 \end{tabular}
 \end{table}
 \end{landscape}

\section{Discussion}
The present study explored a ML-driven approach to classify consciousness among potential LIS patients in order to differentiate them from unconscious individuals based on EEG N400 ERP effects. Detecting consciousness among these patients is challenging because of similarities with disorders of consciousness, a lack of objective biomarkers and a difficult-to-recognize pathogenesis \cite{bruno2011unresponsive,gosseries2014recent,lui2016psychoradiology}. Within the bounds of this experiment, the performance metrics obtained show that an SVM and RF are to different extents, capable of classifying conscious from unconscious individuals. The RF especially achieved high accuracy, sensitivity and specificity when predicting on unseen data, advocating for the further development of a model that uses EEG N400 ERP signals. These results were not unexpected given the success SVMs and RFs have achieved with regards to the classification of consciousness in related contexts \cite{holler2013comparison,engemann2018robust} (see Section \ref{RelatedWork}). Moreover, as has been shown by the robustness check, both models were still capable of differentiating between conscious and unconscious individuals on data from the noun, pre-critical and post-critical sentence regions. The RF produced even better results, possibly because signals acquired during the other regions confounded prediction since they do not actually convey information about the N400 effect. As such, the baselines described in Section \ref{RelatedWork}, up to 40\% related to speech and sound modalities in \citeauthor{steppacher2013n400} (\citeyear{steppacher2013n400}), and 50\% related to passively presented auditory cues in \citeauthor{cruse2014reliability} (\citeyear{cruse2014reliability}), were greatly exceeded by both the SVM and RF, which detected 95\% (95\% in robustness check) and 91\% (100\% in robustness check) of conscious patients based on ERP effects respectively. An explanation for why the present model produced better results is because it was trained on more data of a different stimulus type, used simulated data and used a ML approach to reach classification. These results advocate for the development of a new auxiliary diagnostic tool for the classification of LIS patients based on the ML pipeline presented here. In order to achieve a ML-driven diagnostic tool, some developments should be achieved. The code is openly available on Github \cite{githubrepo} for further development.

\par For one, the number of false positives (conscious classified as unconscious) and false negatives (unconscious classified as conscious) should be reduced to a minimum. Even though every diagnostic tool should pursue high sensitivity to reduce false negatives and high specificity to reduce false positives, diagnostic tools for LIS urge especially for high sensitivity because misdiagnoses generally emanate from false negative classification \cite{zhu2010sensitivity,gosseries2014recent,bruno2011unresponsive}. Therefore, future endeavours should penalize false negatives more heavily than false positives by training the model on more instances of varying nature.

\par The present study ratifies the possibility to create a new auxiliary diagnostic tool for the automatic classification of LIS patients. In order for this possibility to become reality, the presented models should be trained on real data to capture consciousness instead of data that indirectly represents it. Using real data ensures that the performances achieved here can be extended to the actual relationship between ERP effects and (un)consciousness. When similar results are obtained on this data, the models are capable of generalizing to unseen instances, endorsing the development of the diagnostic tool. Even with a relatively low amount of subjects, which is likely given common sampling obstacles faced in LIS \cite{noirhomme2017look}, high sensitivity in the classification of conscious LIS patients can still be achieved by high sensitivity in the classification of unconscious subjects \cite{holler2013comparison}.

\par The consideration should also be made to formalize consciousness by training the models with EEG signals in a resting state so that new data can similarly be collected in resting state. Given that N400 effects can be found in subjects exposed to normatively associated word pairs \cite{cruse2014reliability}, this should not pose any significant obstacles. Further training the data should be based on occurrences where consciousness is formalized by utilizing the consciousness indexes described in \cite{sitt2014large} to ensure validity. This will improve accuracy and generalizability, makes new data collection convenient and reinforces the overall model dexterity. The virtually real-time classification of new instances makes the models presented here promising as an auxiliary diagnostic tool for LIS. If this is done properly, the studies from which the baseline was acquired (see Section \ref{RelatedWork}) can possibly benefit from the presented model as well. The fact that none of these studies used a ML approach makes it interesting to apply the models presented here to those contexts as well, possibly extending the use to more differentiated contexts. Since the studies conducted by \citeauthor{steppacher2013n400} \cite{steppacher2013n400}, \citeauthor{balconi2013disorders} \cite{balconi2013disorders} and \citeauthor{beukema2016hierarchy} \cite{beukema2016hierarchy} are based on different methods to elicited ERP effects, it would be interesting to see how the present models perform in those contexts.

\section{Conclusion}
Concluding, an SVM and RF were trained on N400 ERP effects of data collected by \citeauthor{nicenboim2020words} \cite{nicenboim2020words} in a sentence comprehension task. This data represented conscious individuals by means of EEG signals paired with congruent Cloze scores. Unconscious individuals were simulated based on this dataset, where EEG signals were paired with randomly generated Cloze scores so that a relationship between the two ceased to exist. Both the SVM and RF were able to classify most individuals with a sensitivity of 72\% and 91\% and a specificity of 94\% and 91\% respectively. If the limitations described in the discussion are taken into consideration, results from this study present the opportunity to create a new auxiliary diagnostic tool for the automatic classification of LIS patients among unconscious individuals, so that diagnosis becomes cheap, convenient, fast and reliable, ultimately reducing mortality rates and sustaining optimistic prognosis. Moreover, the model proposes interesting opportunities to be applied to different contexts related to the detection of consciousness.

\newpage
\printbibliography
\newpage

\appendix
\section{Background information}
\label{appendix:bginfo}
Locked-in syndrome, originally classified by Plum and Posner \cite{plum1966diagnosis}, describes a neurological condition in which patients experiences quadriplegia (i.e. complete paralysis of all four limbs), lower cranial nerve paralysis and anarthria (i.e. the total loss of speech), all while retaining consciousness, vertical eye gaze and upper eyelid movement \cite{smith2005locked}. The original definition described mutism instead of anarthria, but this was changed twenty years later since mutism could also imply voluntary unwillingness to speak \cite{patterson1986locked}. Other symptoms include difficulty breathing, dizziness and nausea as well as insomnia \cite{casanova2003locked}. The fundamental difference between LIS and coma and coma-like states lies within the full preservation of consciousness in the former \cite{bauer1979varieties}. These patients thus experience a conscious mind being locked in a physical body, hence the name LIS. Even though consciousness is preserved, attention, executive functions, intellectual ability, perception and visual and verbal memory are often affected \cite{smith2005locked}. Nevertheless, vertical eye movements and blinking allow patients to convey coded communication to their surroundings \cite{bruno2011survey}. Eye movements are generally unaffected because the ocular motor pathways are separate from the affected areas responsible for the syndrome \cite{leon2002review}.

\par The syndrome can arise from central pontine myelinolysis (i.e. the disappearance of neurons and myelin in the pons), as a result of alcoholism, malnourishment or liver diseases \cite{sohn2014locked}. However, the most common causes are ischemic strokes, traumas and brainstem lesions \cite{oken2014brain, desai2015traumatic}. Importantly, early diagnosis has been recognized as a key prospect for functional recovery throughout literature. \citeauthor{sohn2014locked} \cite{sohn2014locked} stated that even though most patients will continue to suffer from neurological deficits, mortality rates are drastically reduced if adequate care is provided. This can be achieved by timely rehabilitation followed by effective and goal-driven hospitalization \cite{casanova2003locked}. \citeauthor{leon2002review} \cite{leon2002review} accentuated that aggressive treatment culminates in functional recovery, which is sanguine for a usually pessimistic prognosis given the high mortality rate of the syndrome.

\par Diagnosis of LIS patients remains however a contemporary challenge. A brainstem lesion is followed by a prolonged comatose period in which patients often gradually become conscious but remain suffering from quadriplegia and anarthria \cite{lesenfants2014independent}. Bedside behavioral examinations are not capable enough of detecting voluntary micro-movements, inherently related to consciousness \cite{gosseries2014recent}, which is why diagnosis is delayed by about 80 days on average \cite{leon2002review}, or completely missed by caretakers and family members \cite{smith2005locked}. The similarities with coma and coma-like states make that LIS patients are easily misdiagnosed without specialized equipment \cite{lesenfants2014independent}. Bedside behavioral examinations can however not be disregarded since suspicion of consciousness can trigger diagnosis in some cases \cite{leon2002review}. The exact number of people suffering from LIS is actually unknown because of these diagnostic obstacles \cite{hachinski2006national}, urging for competent tools that encourage early successful diagnosis.
\newpage

\section{Original data} 
\label{appendix:originaldata}
\citeauthor{nicenboim2020words} \cite{nicenboim2020words} recorded EEG signals for all their subjects from 32 scalp sites using an ANT Neuro amplifier manufactured by TMSi. Eye movements and blinks were recorded with bipolar electrodes placed to monitor both horizontal and vertical electrooculography (EOG). Both EEG and EOG signals were recorded at a 512 Hz sampling rate with a 138 Hz low-pass filter. Recordings were referenced to the left mastoid and later re-referenced to the average of the left and right mastoid channels. Next, recordings were preprocessed with the R \cite{RStudioteam} package \textit{eeguana} \cite{eeguana}, in which data was filtered using a zero-phase band-pass finite impulse response (FIR) filter with pass band-edge frequencies of 0.1 and 30 Hz. This filter, adapted from the Python package MNE \cite{gramfort2013meg}, applied a transition band for low and high edges of 0.10 and 7.50 Hz respectively. An independent component analysis (ICA: \cite{jung2000removing} was used to correct for eye-movements using the deflation-based FastICA algorithm \cite{hyvarinen1997fast,nordhausen2011deflation,miettinen2014deflation}, after which segments containing a voltage difference of over 100 $\mu$V in a time window of 150ms or containing a voltage step of over 50 $\mu$V/ms were rejected. Finally, this signal was segmented and baseline-corrected relative to a 100ms interval preceding the experimental stimulus. The code as well as the scripts \citeauthor{nicenboim2020words} \cite{nicenboim2020words} used to achieve this signal can be found in \cite{nicenboim2019eegdata}. These signals were stored in .RDS files and dependent of the presented sentences, the amount of EEG and EOG signals varied roughly between 30.000 and 900.000 per subject, contingent of the length of their experiment. Besides these EEG and EOG signals, \citeauthor{nicenboim2020words} \cite{nicenboim2020words} included a multitude of features in these .RDS files. These files, referred to as original files, were then preprocessed as described in Section \ref{Data}.
\newpage

\section{ML Pipeline}
\label{Appendix3}
\includegraphics[page=1,scale=0.7]{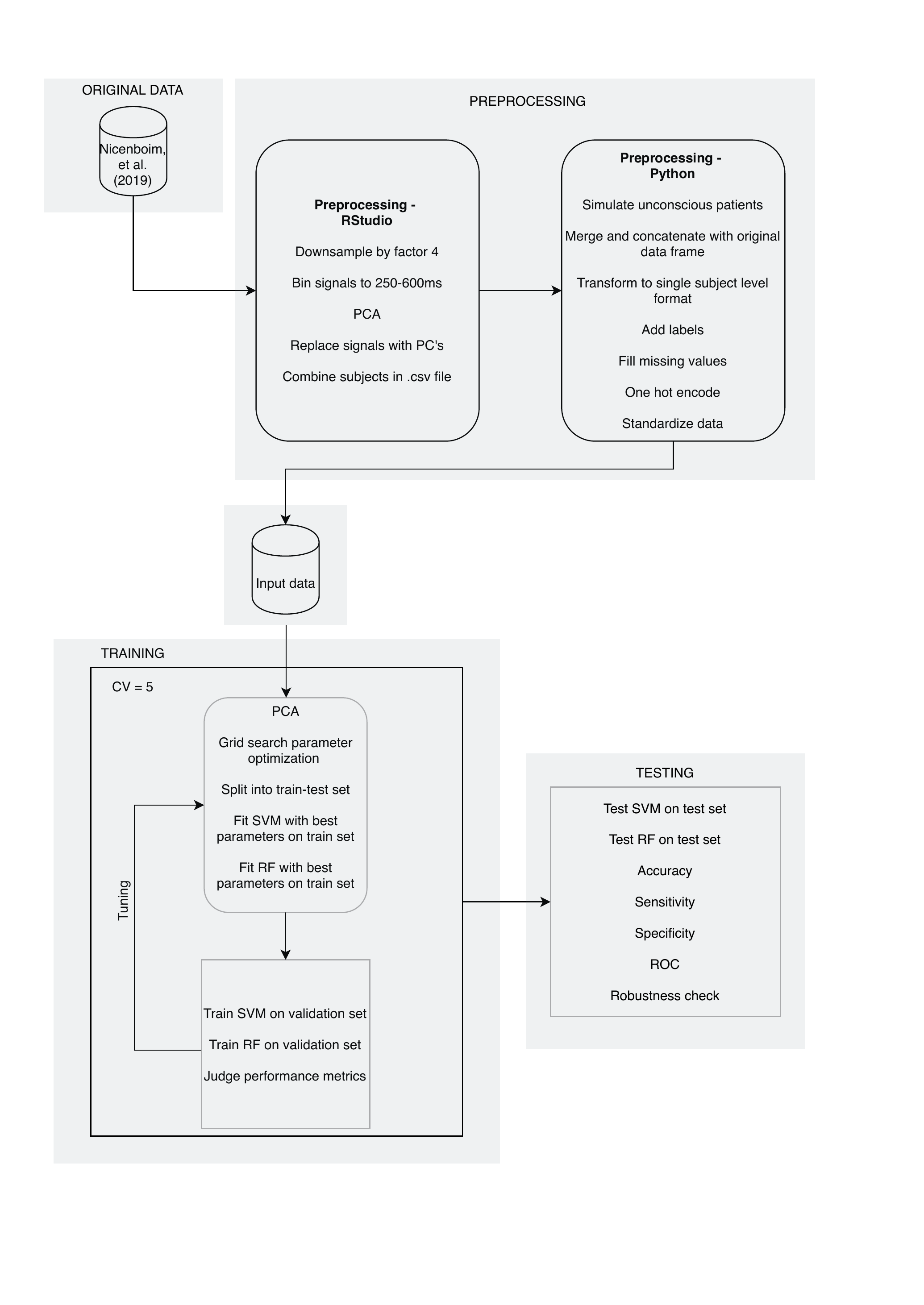}
Preprocessing and analysis pipeline
(\url{https://github.com/DanielvdC/LISclassification/blob/master/Visualizations/ML\%20pipeline.pdf} for full size)

\end{document}